\documentclass[twocolumn,showpacs,preprintnumbers,amsmath,amssymb,prl]{revtex4}

\usepackage{graphicx}
\usepackage{dcolumn}
\usepackage{epsf}

\begin{document}

\title{Phase textures induced by dc current pairbreaking in
multilayer structures and two-gap superconductors.}
\author{A. Gurevich$^{1}$ and V.M. Vinokur$^{2}$.}
\affiliation{$^{1}$National High Magnetic Field Laboratory, Florida
State University, Tallahassee, Florida, 32310 \\
$^{2}$Materials Science Division, Argonne National Laboratory,
Argonne, Illinois, 60439}

\date{\today}

\begin{abstract}
We predict current-induced formation of equilibrium phase textures
for a multicomponent superconducting order parameter. Using the
two-component Ginzburg-Landau and Usadel equations, we show that,
for weakly coupled co-moving superconducting condensates, dc current
I first causes breakdown of the phase locked state at $I>I_{c1}$
followed by the formation of intrinsic phase textures well below the
depairing current $I_d$. These phase textures can manifest
themselves in multilayer structures, atomic Bose condensate mixtures
in optical lattices and two-gap superconductors, particularly
MgB$_2$, where they can result in oscillating and resistive
switching effects.

\end{abstract}
\pacs{\bf 74.20.De, 74.20.Hi, 74.60.-w}

\maketitle

Interest in novel effects caused by multicomponent order parameters
in heavy fermion and organic superconductors \cite{su} has been
recently amplified by the discovery of two-gap superconductivity in
MgB$_2$\cite{mgb2}, which exhibits an anomalous increase of the
upper critical field by impurities \cite{ag} and intrinsic Josephson
effect between two weakly coupled order parameters $\Psi_1=\Delta_1
e^{i\theta_1}$ and $\Psi_2=\Delta_2 e^{i\theta_2}$ in $\sigma$ and
$\pi$ bands. Excitations of the interband phase difference
$\theta=\theta_1-\theta_2$ can be either the phonon-like Legget
modes \cite{legget} or phase textures formed by $2\pi$ interband
phase slips. Phase textures and peculiar vortex properties
\cite{babaev} in two-gap superconductors are generic manifestations
of the multicomponent nature of the order parameter, which have
analogs in superfluid $^3$He \cite{legget} and models of color
superconductivity in the particle physics \cite{colorsc}.

Spontaneous formation of phase textures breaking the time reversal
symmetry is inhibited by interband coupling, which locks the phases
of $\Psi_1=\Delta_1 e^{i\theta_1}$ and $\Psi_2=\Delta_2
e^{i\theta_2}$. However, for weak interband coupling characteristic
of MgB$_2$, the order parameters $\Psi_1$ and $\Psi_2$ can be
unlocked by electric fields. The resulting {\it  non-equilibrium}
charge imbalance generates interband phase textures, which do not
carry magnetic flux and thus do not interact with weak magnetic
fields and supercurrents \cite{gv}. In this Letter we show that, for
sufficiently strong superconducting currents, formation of phase
textures does not necessarily require any nonequilibrium conditions
and can result from dc current pairbreaking, which decouples
$\Psi_1$ and $\Psi_2$ well below the global depairing threshold.
Thus, the phase textures are indeed a rather generic feature of {\it
equilibrium} current-carrying states, which are not specific to
two-gap superconductors, but can be realized in any system with at
least two different co-moving superconducting or Bose condensates,
in either the coordinate or the momentum space. The examples range
from a weakly coupled thin film bilayer (Fig. 1) to a mixture of two
weakly coupled atomic Bose condensates in optical lattices
\cite{bec}. The bilayer in Fig. 1 can be mapped onto a two-gap
superconductor in which two electron bands correspond to the
different films, and the interband coupling corresponds to the
interlayer Josephson energy.

\begin{figure}                  
    \epsfxsize= 0.9\hsize
    \centerline{
    \vbox{
    \epsffile{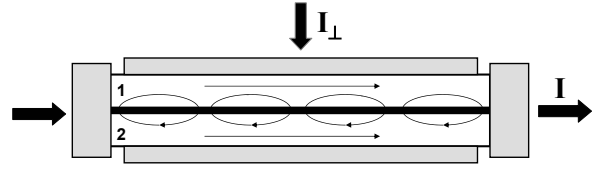}
    }}
    \caption{Formation of phase textures in current-carrying bilayer. Thick black line shows
    the interlayer Josephson contact and gray rectangles show current leads.  }
    \label{Fig.1}
    \end{figure}

The mechanism of texture formation is particularly transparent in a
bilayer, which also reveals peculiar nonstationary effects such as
switching, parametric amplification, and flux flow oscillations
controlled by the applied current in the absence of a dc magnetic
field. The sheet supercurrent $I=I_1+I_2$ in the bilayer is
sustained by the phase gradients $\theta_1^{\prime}$ and
$\theta^{\prime}$ in films 1 and 2, where
$I_1=-cd_1\phi_0\theta_2^{\prime}/8\pi^2\Lambda_1^2$,
$I_2=-cd_2\phi_0\theta_1^{\prime}/8\pi^2\Lambda_1^2$, $\Lambda_1$
and $\Lambda_2$ are the London penetration depths, $\phi_0$ is the
flux quantum, c is the speed of light, and the vector potential
${\bf A}$ is negligible for small film thicknesses $d_1\ll\Lambda_1$
and $d_2\ll\Lambda_2$.  For small current $I$, the minimum energy
corresponds to the phase-locked state with the global phase gradient
$\theta_1^\prime=\theta_2^\prime=Q= -8\pi^2
(\Lambda_1^2/d_1+\Lambda_2^2/d_2)I/c\phi_0$ for which the interlayer
Josephson energy vanishes.

The situation changes at higher $Q\sim 1/\xi_2$, if the coherence
length $\xi_2$ in film 2 is greater than $\xi_1$ in film 1. Then the
Ginzburg-Landau (GL) current pairbreaking suppresses the gap
$\Delta_2(I) = \Delta_2[1-(\xi_2Q)^2]^{1/2}$ more than the other gap
$\Delta_1(I) = \Delta_1[1-(\xi_1Q)^2]^{1/2}$, so the maximum
phase-locked current is limited by the depairing threshold
$Q=1/\sqrt{3}\xi_2$ in film 2. For $Q>1/\sqrt{3}\xi_2$, film 2 goes
normal, forcing current redistribution, which restores
superconductivity in film 2, but causes a gradient of the phase
difference $\theta^\prime(x)$ and the oscillating perpendicular
Josephson current $J_c\sin\theta$ along the bilayer. As a result,
current loops appear in which currents flow parallel to I in the
"stronger" film 1 and antiparallel to $I$ in the "weaker" film 2, as
shown in Fig. 1. Such decoupling transition occurs if the gain in
the condensation energy due to the current redistribution exceeds
the loss in the Josephson energy for a single vortex.

A theory of this transition can be developed using the GL free
energy $\Im=\int F d^2{\bf r}$,
    \begin{eqnarray}        
    F=\sum_md_m\bigl[-\tilde{\alpha}_m\Delta_m^2+\frac{\beta_m}{2}\Delta_m^4
    +\gamma_m(\nabla\Delta_m)^2] \nonumber \\
     +B^2/8\pi +\epsilon_J(1-\cos\theta),\qquad\qquad
    \label{fder}
    \end{eqnarray}
where $m$ labels films 1 and 2,
$\tilde{\alpha}_m=\alpha_m-Q_m^2\gamma_m$, ${\bf
Q}_m=\nabla\theta_m+2\pi{\bf A}/\phi_0$, and ${\bf
B}=\nabla\times{\bf A}$. The last term defines the sheet coupling
energy $\epsilon_J = \phi_0J_c/2\pi c$ where $J_c$ is the Josephson
critical current density. For weak coupling, $\theta(x)$ varies over
the length $L_\theta$ much greater than $\xi_m =
(\gamma_m/\alpha_m)^{1/2}$. Thus,
$\Delta_m^2=\tilde\alpha_m/\beta_m$ to the zero accuracy in
$\epsilon_J$, and the thermodynamic potential G for slowly varying
$Q_m(x)$ and fixed $I$, takes the form:
    \begin{eqnarray}
    G=\int d^2{\bf r}\bigl[-\sum_m\frac{d_m\alpha_m^2}{2\beta_m}(1-Q_m^2\xi_m^2)^2
    \nonumber \\
    +\epsilon_J(1-\cos\theta) + IA/c - \phi_0J_\perp\theta/2\pi
    c\bigr].
    \label{inter}
    \end{eqnarray}
Here $J_\perp$ is the current density injected {\it perpendicular}
to the layers, and the sheet current $I(Q)$ along the bilayer is
determined from $\partial G/\partial A=0$. Next, we extract the
$\theta$-dependent energy $G_\theta$ of phase textures, expressing
$Q_m$ in terms of $\theta$ and $I$ in Eq. (\ref{inter}). For slow
variations of $\theta(x)$, a 
quadratic gradient expansion of $G_\theta$ yields
        \begin{equation}
        G_\theta=\epsilon_J\int[\frac{L_\theta^2}{2}(\nabla\theta)^2
        +1-\cos\theta-\eta\theta-({\bf
        Q}\nabla\theta)h\bigr]d^2{\bf r},
        \label{Gt}          
        \end{equation}
where $\eta=J_\perp/J_c$. The phase length $L_\theta(Q)$, the
coupling parameter $h(Q)$ and $I(Q)$ depend parametrically on the
background gauge-invariant phase gradient Q:
    \begin{eqnarray}
    h=\frac{8\epsilon_1\epsilon_2(\xi_2^2-\xi_1^2)Q^2}{[(1-3\xi_1^2Q^2)\epsilon_1+(1-3\xi_2^2Q^2)\epsilon_2]\epsilon_J}
    \label{be} \\
    L_\theta^2=\frac{4\epsilon_1\epsilon_2(1-3\xi_1^2Q^2)(1-3\xi_2^2Q^2)}
    {[(1-3\xi_1^2Q^2)\epsilon_1+(1-3\xi_2^2Q^2)\epsilon_2]\epsilon_J}
    \label{Lg} \\
    I=-
    Q[\epsilon_1(1-\xi_1^2Q^2)+\epsilon_2(1-\xi_2^2Q^2)]8\pi
    c/\phi_0.
    \label{I}
    \end{eqnarray}
Here $\epsilon_1=d_1\xi_1^2\alpha_1^2/2\beta_1$ and
$\epsilon_2=d_2\xi_2^2\alpha_2^2/2\beta_2$ are characteristic
condensation energies in films 1 and 2 at $I=0$. For weak Josephson
coupling, $\epsilon_J\ll \mbox{min} (\epsilon_m/\xi_m^2)$, the phase
length $L_\theta$ is much greater than $\xi_1$ and $\xi_2$, except
for special cases discussed below. From Eq. (\ref{Gt}) we obtain the
following dynamic equation for $\theta({\bf r},t)$:
    \begin{equation}
    \tau^2\ddot{\theta}+\tau_r\dot{\theta}=
    L_\theta^2\nabla^2\theta-\sin\theta-
    \mbox{div}({\bf Q}h)+\eta,
    \label{tdh}
    \end{equation}
where $\tau^2=C\phi_0/2\pi cJ_c$, $\tau_r=\phi_0/2\pi c RJ_c$, and C
and R are the sheet capacitance and quasiparticle ohmic resistance
of the Josephson contact, respectively.

Eq. (\ref{Gt}) resembles the free energy of a long Josephson contact
in a magnetic field, but in our case the driving term $({\bf
Q}\nabla\theta)h$ results from the pairbreaking asymmetry of the
layers. Here $h(Q)\propto I^2$ at small I, but $h(Q)$ diverges at
the global depairing current $I_d=I(Q_d)$ where
$Q_d^2=(\epsilon_1+\epsilon_2)/3(\epsilon_1\xi_1^2+\epsilon_2\xi_2^2)$.
Yet, $I_d$ cannot be reached because the phase-locked state becomes
unstable above the depairing threshold, $Q>1/\sqrt{3}\xi_2<Q_d$, in
film 2.

At $Q\approx1/\sqrt{3}\xi_2$ the gradient term in Eq. (\ref{Gt})
changes sign so $G_\theta$ should be expanded in higher order
spatial derivatives of $\theta$, which add the stabilizing term
$\ell^2 L_0^2(\nabla^2\theta)^2/2$ into Eq. (\ref{Gt}) where
$L_\theta^2(Q)\approx L_0^2(1-3\xi_2^2Q^2)$,
$L_0=2(\epsilon_2/\epsilon_J)^{1/2}$, and $\ell\sim \xi_2$. In the
critical region, $Q\approx 1/\sqrt{3}\xi_2$, a small perturbation
$\delta\theta = \theta_0\cos kx$ changes the energy by $\delta
G_\theta\propto 1-k^2L_0^2(3\xi_2^2Q^2-1)+\ell^2L_0^2k^4$. Hence,
$\delta G_\theta$ is minimum at $k_m^2=(3\xi_2^2Q^2-1)/2\ell^2$, and
$\delta G_\theta(k_m)<0$ if $3\xi_2^2Q^2>1+2\ell/L_0$. The bilayer
becomes unstable with respect to phase perturbations with the wave
vector $k_m=(\ell L_0)^{-1/2}$ at $Q>Q_{c2}$ where
    \begin{equation}
    Q_{c2}\simeq\frac{1}{\sqrt{3}\xi_2}\left(1+\frac{\ell}{2}\sqrt{\frac{\epsilon_J}{\epsilon_2}}\right),
    \qquad k_m\simeq
    \frac{\epsilon_J^{1/4}}{(2\ell)^{1/2}\epsilon_2^{1/4}}.
    \label{spin}
    \end{equation}
At the spinodal point $Q=Q_{c2}$ the phase-locked state is
absolutely unstable, but stable large-amplitude textures become
energetically favorable at a lower $Q\thickapprox Q_{c1}$ due to
proliferation of interlayer $2\pi$ phase slips $\theta =
4\tan^{-1}\exp(x/L_\theta)$ similar to the Josephson vortices at
$H>H_{c1}$. At $Q=Q_{c1}\ll Q_{c21}$ the energy of a single phase
slip $8\epsilon_JL_\theta$ equals the gain in the condensation
energy $2\pi\epsilon_JQ_{c1} h(Q_{c1})$ in Eq. (\ref{Gt}). Hence,
    \begin{equation}
    Q_{c1}=\left[\frac{(\epsilon_1+\epsilon_2)\epsilon_J}
    {\pi^2\epsilon_1\epsilon_2(\xi_2^2-\xi_1^2)^2}\right]^{1/6},
    \label{jtj}
    \end{equation}
For $I>I_{c1}=8\pi cQ_{c1}(\epsilon_1+\epsilon_2)/\phi_0$, the
minimum energy corresponds to the chain of phase slips spaced by
$a(I)$ for which $\sin\theta/2=\mbox{cn}(x/pL_\theta |p^2)$
\cite{j}. Here $\mbox{cn}(x|p^2)$ is an elliptic function, and
    \begin{equation}
    a=2L_\theta pK(p^2),\qquad I^3 = I_{c1}^3E(p^2)/p,
    \label{aJ}
    \end{equation}
where $K(p^2)$ and $E(p^2)$ are the complete elliptic integrals
defined by the parameter $0<p<1$. The period $a(I)$ diverges
logarithmically at $I\to I_{c1}$, but then $a\approx
0.5\pi^2L_\theta (I_{c1}/I)^3$ decreases rapidly as $I$ further
increases. As a result, the current density in layer 2 remains close
to $J_{2d}$, while $J_1$ increases up to $J_{1d}$, giving the
maximum current $I_{d}(T)=d_1J_{d1}(T)+d_2J_{d2}(T)$ for decoupled
layers. Formation of the phase textures at $Q=Q_{c1}$ is therefore a
first order phase transition with a spinodal decomposition of the
phase-locked state at $Q>Q_{c2}$.

The above GL theory of current-induced decoupling of spatially
separated condensates holds if $T_{c1}$ and $T_{c2}$ are not too
different, for example, in a bilayer made of the same superconductor
with different concentrations of nonmagnetic impurities. It also
implies that $T<T_{c2}<T_{c1}$ so superconducting states in both
layers are weakly coupled, unlike the case $T>T_{c2}$ for which
$\Delta_2$ is induced by proximity effect, and the phase-locked
state persists for all $I<I_d$ \cite{j}. Now we turn to the
condensate decoupling in the momentum space, focusing on interband
phase textures in two-gap superconductors. We use here the
quasiclassic equations of two-gap superconductivity in the dirty
limit \cite{ag}, which also enable us to calculate the phase
textures in bilayers for all $T$:
    \begin{eqnarray}
    \omega f_m-\frac{D_m^{\alpha\beta}}{2}[g_m\Pi_\alpha\Pi_\beta f_m-f_m\nabla_\alpha\nabla_\beta g_m]
    \nonumber \\
    =\Psi_mg_m+\Gamma_{m\overline{m}}(g_{m} f_{\overline{m}}-g_{\overline{m}} f_{m}),\qquad
    \label{uz} \\
    \Psi_m=2\pi T\sum_{\omega >
    0}^{\omega_D}\sum_{m'}\lambda_{mm'}f_{m'}({\bf r}, \omega),
    \label{d} \\
    J^\alpha=-2\pi eT \mbox{Im}\sum_{\omega>0}\sum_{m}N_mD_m^{\alpha\beta}f_{m}^*\Pi_\beta f_{m}.
    \label{j}
    \end{eqnarray}
Here $f_m({\bf r},\omega)$ and $g_m({\bf r},\omega)$ are the Usadel
functions in the m-th band, $|f_{m} |^2+g_{m}^2=1$, $\omega = \pi
T(2n+1)$, $D_m^{\alpha\beta}$ are the intraband diffusivities,
$\overline{m}=2$ if $m = 1$, and $\overline{m}=1$ if $m = 2$,
$\Gamma_{m\overline{m}}$ are the interband scattering rates, ${\bf
\Pi}=\nabla + 2\pi i {\bf A}/\phi_0$, $N_m$ is the partial density
of states, $\lambda_{mm'}$ are the BCS coupling constants, and
$N_1\lambda_{12}=N_2\lambda_{21}$. The indices 1 and 2 correspond to
the $\sigma$ and $\pi$ bands of MgB$_2$ for which
$\lambda_{\sigma\sigma}\approx 3\lambda_{\pi\pi}\approx
8\lambda_{\sigma\pi}$, $N_\pi\approx 1.3 N_\sigma$ \cite{golub}.

Eqs. (\ref{uz})-(\ref{j}) can be obtained by varying the free energy
$\Im =\int F d^3{\bf r}$ where
    \begin{eqnarray}
    F=\frac{1}{2}\sum_{mm^\prime}\Psi_m\Psi_{m^\prime}^*N_m\lambda_{mm^\prime}^{-1}
    +F_1+F_2+F_i +\frac{B^2}{8\pi},
    \label{FF} \\
    F_m=2\pi T\sum_{\omega > 0}N_m[\omega (1-g_{m})-\mbox{Re}(f_{m}^*\Psi_m)
    \nonumber \\
    +D_m^{\alpha\beta}(\Pi_\alpha f_{m}\Pi_\beta^* f_{m}^*+\nabla_\alpha g_{m}\nabla_\beta
    g_{m})/4].\qquad
    \label{Fm}
    \end{eqnarray}
Here $F_m$ is the intraband energy, and $F_i=2\pi
T\mbox{Re}\sum_{\omega
> 0} (N_1\Gamma_{12}+N_2\Gamma_{21})(g_{1} g_{2}+f_{1}^*f_{2}-1)$ is
due to interband scattering. The first term in Eq. (\ref{FF})
contains the Josephson-like interband coupling energy
$-\epsilon_i\cos\theta$ where
$\epsilon_i=\Delta_1\Delta_2\lambda_{12}N_1/w$, and
$w=\lambda_{11}\lambda_{22}-\lambda_{12}\lambda_{21}$.

We derive the equation for the slowly varying ($L_\theta \gg\xi$)
interband phase difference $\theta({\bf r})$, neglecting weak
interband scattering \cite{maz} and expanding Eqs.
(\ref{uz})-(\ref{Fm}) in powers of $Q^2$ and $Q^4$, where the term
$\propto Q^4$ accounts for current pairbreaking in the lowest order
in J. This calculation gives the energy of the phase texture in the
form of Eq. (\ref{Gt}) with $\eta=0$ and
    \begin{equation}
    L_\theta^2=\frac{2\varphi_1\varphi_2}{(\varphi_1+\varphi_2)\epsilon_i},
    \qquad
    h=2Q^2\frac{(\psi_2\varphi_1-\psi_1\varphi_2)}{(\varphi_1+\varphi_2)\epsilon_i},
    \label{hh}
    \end{equation}
where $Q=8\pi^2\Lambda^2J/c\phi_0$,
$\Lambda=\phi_0/[32\pi^3(\varphi_1+\varphi_2)]^{1/2}$ is the London
penetration depth \cite{ag},
    \begin{eqnarray}
    \frac{\varphi_m}{N_mD_m} =\frac{\pi \Delta_m}{8}\mbox{tanh}\frac{\Delta_m}{2T},\qquad\qquad
    \label{phi}\\
    \frac{\psi_m}{N_mD_m^2}=u_m+
    \frac{\pi^2}{32v_m}\bigl[\mbox{tanh}\frac{\Delta_m}{2T}+
    \frac{\Delta_m}{2T}\mbox{sech}^2\frac{\Delta_m}{2T}\bigr]^2.
    \label{psi}
    \end{eqnarray}
Here $u_m=\pi T
\Delta_m^2\sum_{\omega>0}\omega^2/(\omega^2+\Delta_m^2)^{5/2}$,
$v_m=2\pi T\Delta_m^2\sum_{\omega>0}(\omega^2+\Delta_m^2)^{-3/2}$.
Eqs. (\ref{hh})-(\ref{psi}) can also be applied to bilayers, by
replacing $J\to I$, $\epsilon_i\to \phi_0J_c/2\pi c$, and $N_m\to
d_mN_m$. Eq. (\ref{hh}) reduces to Eqs. (\ref{be})-(\ref{I}) near
$T_c$ if $Q\xi_2\ll 1$. However, Eq. (\ref{hh}) also describes the
case (particularly relevant to MgB$_2$) for which $\Psi_1$ and
$\Psi_2$ become weakly coupled only at low T as intraband pairing
causes the Cooper instability in band 2. The simpler GL theory gives
the full dependencies of $L_\theta$ and $h$ on current, while in the
Usadel approach only the main quadratic term in $h(Q)$ can be
obtained analytically for all T. In particular, for $T=0$, Eq.
(\ref{hh}) yields
    \begin{eqnarray}
    L_\theta^2 = \frac{\pi N_1N_2\Delta_1^2\Delta_2^2\xi_1^2\xi_2^2}
    {4(N_1\Delta_1^2\xi_2^2+N_2\Delta_2^2\xi_1^2)\epsilon_i}, \qquad
    \label{lo} \\
    J_{c1}=\frac{kc\phi_0}{8\pi^2\Lambda^2[(\xi_2^2-\xi_1^2)L_\theta]^{1/3}}
    \label{jto}
    \end{eqnarray}
Here the band decoupling current density $J_{c1}$ is defined as
before by $8L_\theta = 2\pi Q_{c1}h(Q_{c1})$ where
$h=(\pi/4+4/3\pi)L_\theta^2(\xi_2^2-\xi_1^2)Q^3$,
$Q_{c1}=8\pi^2\Lambda^2J_{c1}/c\phi_0$, $\xi_m=(D_m/\Delta_m)^{1/2}$
is the intraband coherence length, and
$k=(1/3+\pi^2/16)^{-1/3}\approx 1.017$. If expressed in terms of the
sheet condensation energy densities
$\epsilon_m=N_m\Delta_m^2\xi_m^2/2$, Eqs. (\ref{lo}) and (\ref{jto})
reduce Eqs. (\ref{Lg}) and (\ref{jtj}) to the accuracy of
coefficients $\sim 1$.

The results presented above indicate that the previous calculations
of $J_d$ \cite{jd,carb} for two-gap superconductors in which both
bands were assumed to be phase-locked are only valid at higher
temperatures $T>T_2\sim T_c\exp(1/\lambda -1/\lambda_{22})$ for
which $\Delta_2$ is induced by interband coupling. Here $\lambda =
[\lambda_+ + (\lambda_-^2+4\lambda_{12}\lambda_{21})^{1/2}]/2w$ and
$\lambda_\pm=\lambda_{11}\pm\lambda_{22}$. To estimate $T_2$ for
MgB$_2$, we take $T_c =$40K, $\lambda_{11}=0.8$, $\lambda_{22}=0.3$,
$\lambda_{12}=0.12$, $\lambda_{21}=0.09$ \cite{golub} for which
$T_2\simeq 0.12T_c\simeq$ 5K. This qualitative interpretation is
consistent with the Usadel calculations \cite{jd} and the strong
coupling Eliashberg theory \cite{carb}, which predict a two-hump
$J(Q)$ at low T, which turns into the conventional dome-like $J(Q)$
at higher T. For $T>T_2$, current pairbreaking thus occurs in the
phase-locked state, for which $J_d=\mbox{max} (J(Q))$.

A two-hump $J(Q)$ at low T means that four possible phase gradients
can provide the same current density $J=J(Q)$. Since only states
with $d|J|/dQ>0$ are stable, the two-hump $J(Q)$ would indicate
formation of a stratified flow comprised of parallel channels with
two different phase gradients, similar to the Gunn instability in
semiconductors. However, this stratification is preceded by
interband decoupling, since for $\lambda_{12}\ll\lambda_{22}$, the
band 2 cannot sustain the same Q as the band 1. Thus, interband
current redistribution provided by the phase textures {\it
increases} the GL depairing current density $J_d(T)$ for which
$Q_1=1/\sqrt{3}\xi_1$ and $Q_2=1/\sqrt{3}\xi_2$ as compared to its
phase-locked value ${\tilde J}_d=\mbox{max}J(Q)$:
    \begin{eqnarray}
    \tilde{J}_d=\frac{8\pi
    c(\alpha_1\gamma_1/\beta_1+\alpha_2\gamma_2/\beta_2)^{3/2}}{3\sqrt{3}\phi_0(\gamma_1^2/\beta_1+\gamma_2^2/\beta_2)^{1/2}},
    \label{jd1}\\
    \qquad J_d=\frac{8\pi
    c}{3\sqrt{3}\phi_0}\left(\frac{\alpha_1\sqrt{\alpha_1\gamma_1}}{\beta_1}
    +\frac{\alpha_2\sqrt{\alpha_2\gamma_2}}{\beta_2}\right),
    \label{jd}
    \end{eqnarray}
where $\alpha_m$, $\beta_m$ and $\gamma_m$ are the 2-gap GL
expansion coefficients. The enhancement of $J_d$ is most pronounced
in the case of a clean $\sigma$ band (large $\gamma_1$ \cite{ag}) at
low $T<T_2$ where $J_d(T)$ could be measured by a pulse technique
\cite{kunchur}.

\begin{figure}                  
    \epsfxsize= 0.7\hsize
    \centerline{
    \vbox{
    \epsffile{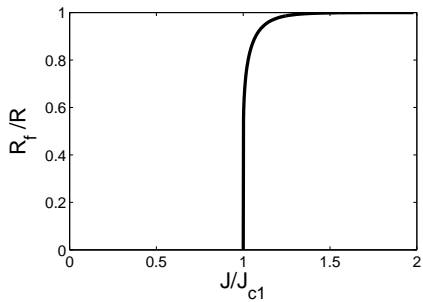}
    }}
    \caption{Flux flow resistance R(I) for the moving phase texture as a function of the
    current I along the bilayer. }
    \label{Fig.2}
    \end{figure}

Phase textures in bilayers manifest themselves in dc transport if
both parallel and perpendicular currents $I$ and $I_\perp$ are
applied, as shown in Fig. 1. For $I<I_{c1}$, no voltage across the
bilayer occurs, but for $I>I_{c1}$, the oscillating voltage $V(x,t)=
-(\phi_0v/\pi cpL_\theta)\mbox{dn}[(x-vt)/pL_\theta |p^2]$ and the
ohmic average voltage $\overline{V}=R_fI_\perp$ appear due to the
phase slip structure moving with the velocity $v(I,I_\perp)$. Here
the resistance $R_f$ is similar to the flux flow resistance of a
long Josephson contact \cite{ff}
    \begin{equation}
    R_f=\pi^2R/4K(p^2)E(p^2),
    \label{rf}
    \end{equation}
where R is the resistance of the interlayer contact, and $p(I)$ is
defined by Eqs. (\ref{aJ}). The dependence $R_f$ on the longitudinal
current I shown in Fig. 2 describes switching between
superconducting and resistive states across the bilayer. Other
effects include a parametric resonance caused by superimposed ac
currents $I_\perp(t)$ and $I(t)$, since $I(t)$ modulates the
parameters in Eq. (\ref{tdh}). Therefore, the geometry in Fig. 1 can
provide switching and Josephson flux flow oscillator in
current-operated devices.

Phase textures in two-gap superconductors can move due to {\it
interband} currents produced, for example, by nonequilibrium charge
imbalance \cite{gv}. One could also expect a kink in the nonlinear
low-frequency $(\omega\ll\Delta_2)$ rf surface resistance as the
field amplitude $H_0$ exceeds the onset of the interband phase slip
formation $H_\theta = 4\pi J_{c1}\Lambda/c$. To estimate $H_\theta$
for MgB$_2$, we assume that the interband breakdown occurs as the
screening current density $J(0)\simeq cH_0/4\pi\Lambda$ on the
surface produces the phase gradient $Q=8\pi^2\Lambda^2J(0)/c\phi_0$
exceeding the GL depairing limit $Q_{c2}=1/\sqrt{3}\xi_\pi$ in $\pi$
band. Hence, $H_\theta = \phi_0/2\sqrt{3}\Lambda\xi_\pi\sim
H_c\xi_\sigma/\xi_\pi$, where $H_c$ is the thermodynamic critical
field, and the ratio $\xi_\sigma/\xi_\pi$ can be strongly affected
by impurities \cite{ag}. Taking $\xi_\sigma/\xi_\pi \approx 0.3$ for
MgB$_2$ single crystals \cite{esk} and $H_c(0)\simeq 0.3$T, we find
$H_\theta(0)\sim 0.1$T. Such fields cause breakdown of the linear
London electrodynamics, affect properties of vortex lattice,
penetration vortices through surface barrier, etc.

In conclusion, two weakly coupled co-moving superconducting
condensates can undergo a first order phase transition into a phase
textured state well below the global depairing current. Such
textures controlled by current result in resistive switching and
oscillating effects.

This work was supported by US DOE Office of Science under contract
No. W31-109-ENG-38.

{\it Note added.} After this paper was submitted, a phase textured
state with two different winding numbers in weakly coupled Al rings
has been observed \cite{kam}.


\end{document}